\documentclass[twocolumn,showpacs,preprintnumbers,amsmath,amssymb,prd,superscriptaddress]{revtex4-1}

\usepackage{amsmath,amssymb,amsthm,color}
\usepackage{graphicx}
\usepackage{dcolumn}

\newcommand {\beq}{\begin{eqnarray}}
\newcommand {\eeq}{\end{eqnarray}}
\newcommand\be{\begin{equation}}
\newcommand\ba{\begin{eqnarray}}
\newcommand\ea{\end{eqnarray}}
\newcommand\ee{\end{equation}}

\begin{document}

\preprint{CALT-68-2855, IPMU13-0164}

\title{Out of Equilibrium Temperature from Holography}

\author{Shin Nakamura}
\affiliation{Department of Physics, Nagoya University, Nagoya 464-8602, Japan}
\author{Hirosi Ooguri}
\affiliation{California Institute of Technology, 452-48, Pasadena, CA 91125, USA}
\affiliation{Kavli Institute for the Physics and Mathematics of the Universe, WPI Initiative, \\ University of Tokyo, Kashiwa 277-8583, Japan}

\date{\today}
\begin{abstract} 
We define an effective temperature and study its properties for a class of out-of-equilibrium steady states in a heat bath. Our analysis is based on the anti-de Sitter spacetime/conformal field theory (AdS/CFT) correspondence, and examples include systems driven by applied electric fields and branes dragged in plasmas. We found that the effective temperature can be lower than that of the heat bath and that the out-of-equilibrium noise can be smaller than that in equilibrium. We show that a generalization of the fluctuation-dissipation relation holds for the effective temperature. In particular, we generalize the Johnson-Nyquist relation for large electric field. 

\end{abstract}


\maketitle


\noindent
{\emph{Introduction}} --- 
Unlike equilibrium systems, which are under powerful constraints of thermodynamics, only few general properties are
known for systems driven far out of equilibrium. In this paper, we use the holographic principle to define
an effective temperature for a class of nonequilibrium steady states and study its properties.
We consider branes driven by external energy sources and  put them in a heat bath.
The heat bath is represented by a bulk geometry with an event horizon.  The Hawking temperature at the horizon is
identified as the temperature $T$ of the heat bath.

Before we turn on the external energy sources, temperature on the brane is
 the same as that of the bulk.  Holographically, this can be seen by
the fact that the induced metric on the brane
has an event horizon 
with the same Hawking temperature $T$. 

We then pump energy into the brane by turning on an external electric 
field or by dragging the brane in the heat bath. Though this would drive
the brane-bulk system out of equilibrium, it can still be stationary 
if we allow the excess energy on the brane to be released into the bulk. 
In general, different degrees of freedom on the brane observe 
different effective metrics. We find, however, that all these metrics 
have an event horizon at the same location with the same Hawking 
temperature $T_*$. Since the system is not in equilibrium, 
$T_*$ is in general different from $T$ in the heat bath.

The effective temperature $T_*$ on out-of-equilibrium branes has been
computed in some examples. A fundamental string dragged in a heat bath of D3 branes 
has been investigated extensively for applications in quark gluon plasma
\cite{Gubser:2006nz,CasalderreySolana:2007qw}.
Branes with an applied electric field were studied in \cite{Kim:2011qh, Sonner}.
A study on Dp branes rotating in an internal sphere direction is found in \cite{Das:2010yw}.
See also a review \cite{Hubeny:2010ry} and references therein.

However, questions have been raised on physical meaning of $T_*$.
In this paper, we will address these questions and provide a uniform view
on the notion of the effective temperature defined for these nonequilibrium branes. 

For example, it was pointed out in \cite{Giecold:2009cg} that, for
the dragged fundamental string, momentum fluctuations are not
isotropic, indicating deferent temperatures for different degrees of freedom.
We will show that, if we take into account the nonlinear relation between 
the momentum and the velocity in the relativistic setup, the energy distribution is in fact isotropic and obeys the Maxwell-Boltzmann rule at temperature $T_*$.

For branes driven by applied electric fields, our results generalize
that of \cite{Kim:2011qh, Sonner} in several ways.
We consider a larger class of brane configurations and
we show that the effective temperatures are the same for all degrees of freedom on the brane in each case.  
We also propose a generalization of Johnson-Nyquist relation in the far-from-equilibrium regime.

Surprisingly, we find that $T_* < T$ in certain cases. Namely, pumping energy into the probe
brane turns its temperature lower than that in the bulk.

We would like to note that, 
since our system is open and driven by an external
energy source, the second law of thermodynamics does not necessarily 
apply and $T_* < T$ does not contradict with known general properties of nonequilibrium systems. 
In fact, there is 
a statistical model where $T_{*}$ is less than $T$ \cite{Sasaki-Amari}. Our results show that 
this phenomenon is robust and takes place in a large class of examples. 

While the effective temperature $T_*$ is the same for all the degrees of freedom on the brane,
their fluctuations and dissipations can be different. 
In particular, we find that, in certain cases, the noise for some degrees of freedom can be 
smaller than that in equilibrium, namely the excess noise generated by the driving force can be negative.

In the following, we define and compute the effective temperature for systems driven by electric fields and
for branes dragged in plasmas. We then discuss the fluctuation-dissipation relation
and other properties of the temperature and the noise around the nonequilibrium steady states. 
We compare our definition of the effective temperature with those defined in the literature on 
nonequilibrium statistical physics such as \cite{effectiveT}. We also discuss to what extent our definition
applies to a more general class of nonequilibrium systems. 

\bigskip

\noindent
{\emph{Electric Field}} --- Consider a quantum field theory in 
$(p+1)$-dimensions at temperature $T$, 
which has a holographic gravity description with the metric,
\begin{equation}
  ds^2 = g_{tt} dt^2 + g_{xx} \sum_{i=1}^p (dx^i)^2
+ g_{rr} dr^2 +  g_{\theta\theta} d\Omega^2,
\label{metric}
\end{equation}
where $t$ is the time coordinate,
$x^i$ are the spatial coordinates for the $(p+1)$-dimensional theory, 
and $d\Omega^2$ is a metric on some compact space, which reflects 
symmetry and other properties of the theory.
We assume that the bulk geometry has a boundary at $r = \infty$ and
that the metric components depend only on $r$. Typically, $g_{tt}$ and $g_{xx}$ diverge and
$g_{rr}$ vanishes at the boundary. 
The dilaton in the bulk (\ref{metric}) may also depend on $r$. 
We use this bulk geometry as a heat bath and assume that the metric has an event horizon with temperature $T$. 

We then introduce a $(q+1)$-dimensional defect in the story, realized
as a probe $(q+1+n)$-brane, which extends in $(q+1)$ dimensions along the boundary
and in the $r$ direction, and is wrapped on an $n$-dimensional subspace of the compact space 
represented by $d\Omega^2$ in the target space metric. 
We assume that the low energy effective
theory is described by the Dirac-Born-Infeld (DBI) action,
\begin{equation}
 {\cal L} = e^{-\phi} \sqrt{- {\rm det} G },
\end{equation}
where 
\begin{equation}
G_{ab} = \partial_a X^\mu \partial_b X^\nu g_{\mu\nu} + F_{ab},
~~ (a,b =0,1,...,q+1),
\end{equation}
$X^\mu$ are the embedding coordinates (we ignore motion of the brane in 
the compact directions), $F_{ab}$ is a Maxwell field strength, and
$e^{-\phi}$ is the dilaton factor times 
the volume of the compact space. 

In certain cases, Wess-Zumino type terms can be generated in the effective action on the brane.
As shown in \cite{CSone,CStwo}, depending on the coefficients of such terms and
the strength of the electric field, they can cause instability 
in the presence of a background electric field. 
We found that, even though Wess-Zumino terms modify dispersion relations of the gauge
and scalar fields on the brane, when they do not cause instabilities,  
they do not modify properties of the effective horizon on the worldvolume and 
in particular the value of the effective temperature $T_*$. 
We plan to discuss these phenomena in a separate paper \cite{inprogress}.

Before we proceed, let us set up some notations. We call 
the symmetric and the antisymmetric parts of $G^{ab}$ (the inverse of $G_{ab}$)
as $G_{(S)}^{ab}$ and $G_{(A)}^{ab}$. We can then write the equations of motion as
\begin{equation}
  \partial_a\left( G_{(S)}^{ab}\partial_b X^\mu  g_{\mu\nu}{\cal L} \right) =0, ~~~
\partial_a\left( G_{(A)}^{ab}{\cal L} \right) =0.
\end{equation}
The inverses of  $G_{(S)}^{ab}$ and $G_{(A)}^{ab}$ are denoted by  $G^{(S)}_{ab}$ and $G^{(A)}_{ab}$
(note that these are not symmetric and antisymmetric parts of $G_{ab}$). 

We consider turning on a constant electric field $E = -F_{01}$ in the 1 direction along the boundary $r = \infty$. 
If we assume that only $F_{01}$ and $F_{r1}$ are turned on, that they are time independent, and 
that they depend only on $r$, the Bianchi identity shows that $F_{01}$ is in fact constant on the brane. Thus,
we can write,
\begin{equation}
 A_1 = - Et + h(r).
\label{electricpotential}
\end{equation}
The first integral $\partial {\cal L}/\partial F_{r1}$ is constant by the equations of motion,
and we denote the constant by $J$ since it can be interpreted as the current density generated by the electric field $E$. 
The magnetic field $F_{r1} = h'(r)$ on the brane is then expressed as
\begin{equation}
 (F_{r1})^2 = J^2 \frac{g_{rr}}{|g_{tt}|} \frac{ E^2 - |g_{tt}| g_{xx}}{J^2- e^{-2\phi}|g_{tt}| g_{xx}^{q-1}}.
\end{equation}

In the holographic models we consider in this paper, $ |g_{tt}| g_{xx}$ diverges at $r=\infty$ and vanishes at the horizon
$r=r_0$. Thus, $(E^2 - |g_{tt}| g_{xx})$ must vanish somewhere between $r_0$ and $\infty$. Let us call the largest of
zero points as $r_*$. Since $(F_{r1})^2$ must
be nonnegative, the brane configuration becomes unphysical for $r < r_*$ unless $(J^2 - e^{-2\phi}|g_{tt}| g_{xx}^{q-1})$
also vanishes at $r_*$. Since $r_*$ defined as the largest solution of $E^2 = |g_{tt}|g_{xx}(r_*)$, it is a function of $E$
and so is $J^2 = e^{-2\phi}|g_{tt}| g_{xx}^{q-1}(r_*)$. The resulting relation between the current density $J$ and the electric field $E$
determines the nonlinear conductivity in this nonequilibrium setup~\cite{Karch:2007pd}. Using this method,
negative differential conductivity and associated
nonequilibrium phase transitions have been 
derived \cite{Nakamura:2010zd, Nakamura:2012ae}. 

The point $r_*$ turns out to be the location of the horizon with respect to an effective metric on the brane~\cite{Kim:2011qh,Kim:2011zd}. To see this,
we look at fluctuations of the embedding coordinate: $X^\mu + \delta X^\mu$. The linearized equations of motion
for $\delta X^\mu$ takes the form,
\begin{equation}
 \partial_a \left[ \chi^{-q/2} \sqrt{-{\rm det} G^{(S)}} G^{ab}_{(S)} \partial_b\delta X^\mu\right]  = 0,
\end{equation}
where $\chi(r)$ is some combination of metric components and $e^{-\phi}$ that remains positive for the range of $r$ we are interested in. 
Thus, the effective metric on the brane experienced by fluctuations $\delta X^\mu$ is $\tilde{g}_{ab} = \chi^{-1} G^{(S)}_{ab}$.

We can diagonalize the effective metric by 
introducing a new time coordinate $\tau$ defined by $d\tau = dt + \frac{\tilde{g}_{tr}}{\tilde{g}_{tt}}dr$ as, 
\begin{equation}
 ds^2_{{\rm eff}} = - \frac{|g_{tt}|g_{xx} - E^2}{\chi g_{xx}} d\tau^2 + 
\frac{e^{-2\phi}g_{xx}^{q-1}|g_{tt}| g_{rr}/\chi}{e^{-2\phi}|g_{tt}| g_{xx}^{q-1}-J^2} dr^2 + \cdots.
\end{equation}
In particular, 
\begin{equation}
ds^2 = - a (r-r_*) d\tau^2 +  \frac{b}{r-r_*}dr^2  + \cdots,
\end{equation}
near $r = r_*$ for some $a$ and $b$. This shows that there is a horizon at $r=r_*$ on the worldvolume. 
The Hawking temperature is computed as,
\begin{equation}
 T_* = \frac{1}{4\pi}\sqrt{\frac{a}{b}}
=\frac{1}{4\pi}\left. \sqrt{\frac{(g_{tt}g_{xx})'(e^{-2\phi}g_{tt}g_{xx}^{q-1})'}{e^{-2\phi} g_{xx}^q |g_{tt}|g_{rr}}}\right |_{r=r_*},
\label{Hawking}
\end{equation}
where $^{\prime}$ denotes the derivative with respect to $r$.

It turns out that the gauge field on the brane also observes the same Hawking temperature. The linearized
equations for gauge field fluctuations $\delta f_{ab}$ are
\begin{eqnarray}
\partial_{b}\left(h^{ab}_{(1)}+h^{ab}_{(2)}\right)=0,
\end{eqnarray}
where 
\begin{eqnarray}
h^{ab}_{(1)}
&=&e^{-\phi}\sqrt{-\det G}
\left[-G_{(S)}^{ac}\delta f_{cd}G_{(S)}^{db}\right],
\\
h^{ab}_{(2)}
&=&e^{-\phi}\sqrt{-\det G}
\left[\frac{1}{2}G_{(A)}^{cd}\delta f_{cd}G_{(A)}^{ab}-G_{(A)}^{ac}\delta f_{cd}G_{(A)}^{db}\right].
\nonumber \\
\end{eqnarray}
We can show that $h^{ab}_{(2)}$ vanishes and that the equation of motion for $\delta f_{ab}$ is equivalent to
the linearized Maxwell equation with the effective metric $\xi^{-1}G_{ab}^{(S)}$, where $\xi(r)$ is some combination 
of metric components and  $e^{-\phi}$ and is in general different from $\chi$ (this has
also been pointed out in \cite{Kim:2011zd}). Since the overall normalization of the metric does not affect the Hawking temperature, 
fluctuations of the gauge field experience the same temperature as that is given in (\ref{Hawking}). 

As an example, consider the heat bath to be $N$ D$p$ branes (with $p<7$) at temperature $T$. Its holographic dual has the metric~\cite{Itzhaki:1998dd},
\begin{eqnarray}
 ds^2 &=& r^{\frac{7-p}{2}}\left[ -(1 - r_0^{7-p}/r^{7-p})dt^2+d\vec{x}^2\right]    
 \nonumber \\
&&+ \frac{dr^2}{r^{\frac{7-p}{2}}\left( 1 - r_0^{7-p}/r^{7-p}\right)} + r^{\frac{p-3}{2}} d\Omega^2,
\label{Dbranemetric}
\end{eqnarray}
where $d\Omega^2$ is the metric on the unit $S^{8-p}$, and the dilaton $e^{\Phi} = e^{\phi_0} r^{(p-3)(7-p)/4}$. 
The horizon is at $r=r_0$ and the Hawking temperature $T$ is given by $c_{0}^{-1}r_0^{\frac{5-p}{2}}$, where $c_{0}=\frac{4\pi}{7-p}$. 

Consider a single D$(q+1+n)$ brane as a probe in this geometry. 
It is extended in $(q+1)$ dimensions along the boundary and stretches in the $r$ direction. 
It is also wrapped on an equatorial $S^{n}$ subspace of the $S^{8-p}$ in such a way that the induced metric of the probe brane in the static gauge agrees with that of the background geometry.
By applying the procedure in the above, we find that the expectation value of the
current $J$ in response to the constant electric field is
\begin{equation}
 J = E \left[r_{0}^{7-p} + E^2\right]^{\frac{C-1}{2}} V_{n}e^{-\phi_0},
\label{J}
\end{equation}
where $V_{n}$ is the volume of the unit $S^{n}$ and
\begin{equation}
   C = \frac{1}{2}\left( q+3 - p + \frac{p-3}{7-p}n  \right). 
   \label{C}
\end{equation}
The horizon $r_*$ for the worldvolume metric is given by
\begin{equation}
 r_* = \left[r_{0}^{7-p} + E^2\right]^{\frac{1}{7-p}}.
\label{horizonlocation}
\end{equation}
Note that the area element of the horizon increases as we turn on the electric field $E$.
The effective Hawking temperature (\ref{Hawking}) in this case is given by, 
\begin{equation}
T_* = c_0^{-1} \frac{\left[\left(c_{0}T\right)^{\frac{14-2p}{5-p}} + C E^2\right]^{\frac{1}{2}}}
{\left[\left(c_{0}T\right)^{\frac{14-2p}{5-p}} + E^2\right]^{\frac{1}{7-p}}}.
\label{branetemp}
\end{equation}
As expected, when we turn off the electric field, the worldvolume horizon $r_*$ approaches the bulk horizon $r_0$ and 
the Hawking temperature $T_*$ reduces to the bulk temperature $T$.

The case at $p=5$ is special since the Hawking temperature $T$ is independent of $r_0$. In this case,  
$(c_{0}T)^{\frac{14-2p}{5-p}}$ in (\ref{branetemp}) should be replaced by $r_{0}^{7-p}$.
When $p<5$, on the other hand, we can adjust the temperature $T$. It is interesting to note
that, even if we take the zero temperature limit $T \rightarrow 0$ in the bulk, 
the effective temperature $T_*$ on the brane does not vanish,
\begin{equation}
  T_* \rightarrow c_0^{-1} C^{1/2} E^{\frac{5-p}{7-p}},~~~(T \rightarrow 0),
 \end{equation}
 assuming that $C$ is positive. We should note that the only case with $C < 0$ that can be realized
as a straightforward intersecting brane configuration in string theory is a D6-D2 system, but it is not
clear if we can use $N$ D6 branes at finite temperature as a heat bath since it has a negative specific heat. 
Otherwise $C$ is always positive and we can take a smooth $T \rightarrow 0$ limit as in the above. 

Let us examine this formula for $T_*$ for some examples. When $p=3$, namely, when the heat bath is made of $N$ D3 branes, we have 
$r_* = (r_0^4 + E^2)^{1/4}$ and 
\begin{equation}
T_* = \pi^{-1} \frac{\left[(\pi T)^4 + \frac{q}{2} E^2\right]^{\frac{1}{2}}}{[(\pi T)^4 + E^2]^{\frac{1}{4}}}.
\end{equation}
Note that,  for 
$q \geq 1$,  $T_*$ is an increasing function of $E^2$. Since we need at least one spatial dimension on the defect to turn on
the electric field, this covers all the cases relevant for $p=3$. 
Thus, in this case, pumping energy into the probe brane raises its temperature from $T$ to $T_*$.
In particular, when $q=2$, we find $T_* = \pi^{-1}[(\pi T)^4+E^2]^{\frac{1}{4}}$, reproducing the result by \cite{Sonner}.

Surprisingly, this is not always the case in higher dimensions.
More explicitly, 
\begin{eqnarray}
T_{*}=T+\frac{1}{2}\left(C-\frac{2}{7-p} \right)\frac{E^{2}}{(c_0T)^{\frac{14-2p}{5-p}}}T+O(E^{4}),\:\:
\end{eqnarray}
and $T_{*}$ is lower than $T$ at the order of $E^2$ if
\begin{eqnarray}
C-\frac{2}{7-p}= \frac{1}{2}\left( q + 3 - p + \frac{(p-3)n-4}{7-p}  \right) < 0.
\end{eqnarray}
Note that $r_{*}>r_{0}$ does not imply $T_{*}>T$.

For example, when $n=0$ and $p=4$, we have
\begin{equation}
T_* = \frac{1}{c_{0}}\frac{[(c_{0}T)^6 + \frac{q-1}{2}E^2]^{\frac{1}{2}}}{[(c_{0}T)^6 +E^2]^{\frac{1}{3}}}
\end{equation}
with $c_{0}=4\pi/3$, and this is a decreasing function of $E^2$ for $q=1, 2$.  The case with $q=1$ can be realized as the
worldvolume theory of a single D$2$ brane probing D$4$ branes. In this case, 
\begin{equation}
  T_* = \frac{c_{0}^2 T^3}{[(c_{0}T)^6 + E^2]^{\frac{1}{3}}},
\end{equation}
showing that it is a decreasing function of the electric field. 
Thus, we can {\it lower} the temperature on the D2 brane in the heat bath of D4 branes 
by turning on the electric field. 

More generally, $C$ is an increasing function of $q$ for fixed $p$ and $n$. 
This means that $T_*$ tends to be lower when the codimensions $(p-q)$ of the defect is larger. 
Note that this effect is of the order of $E^{2}$ and is beyond the linear response regime.

It is interesting to note that, although the effective temperature $T_*$ decreases in some of the cases discussed here, 
the area of the effective horizon set by $r_*$ computed in (\ref{horizonlocation}) is always greater than 
the area of the horizon in the bulk. 
If one could define an entropy on the brane and if it is related to the area of the horizon, 
the increase of the entropy under decreasing temperature would imply negative specific heat. It is not clear, however, 
if one can define an entropy in the far-out-of equilibrium setup discussed here. 
More importantly, we have analyzed linear perturbations to the brane configurations---this is how we computed the effective temperature---and found no indications of instability. 
In particular, frequencies of linear perturbation modes are all real-valued.

\bigskip
\noindent
{\emph{Drag Force}} --- We can use a similar method to compute an effective temperature
on a probe brane dragged by an external force in a heat bath. Consider a D$(q+1+n)$ brane as a probe
in a heat bath of D$p$ branes. We assume that the probe brane is wrapped on an equatorial $S^{n}$ part of the $S^{8-p}$
and extended in the $r$ direction. This represents a $q$-dimensional object in the $(p+1)$-dimensional
theory on the D$p$ branes. Choose a spatial coordinate $X$ in $(p+1)$ dimensions transverse to 
the probe brane and apply the ansatz,
\begin{equation}
 X = vt + x(r), 
\label{branebending}
\end{equation}
analogous to (\ref{electricpotential}) in the previous case. We can interpret $v$ as the velocity of the 
probe brane. The equations of motion for $x(r)$ can
be solved by using the first integral $\partial {\cal L}/\partial x'$ \cite{Herzog:2006gh,Gubser:2006bz,Karch}. 

We can then compute the
effective metric on the probe brane experienced by fluctuations of the embedding coordinates and find
that it has a horizon $r_*$ at
\begin{equation}
   r_* = \frac{r_0}{(1-v^2)^{\frac{1}{7-p}}}.
\end{equation}
As in the previous case, we find the area element of the horizon increases as a function of the velocity $v$. 
The Hawking temperature $T_*$ for the worldvolume metric is,
\begin{equation}
 T_* = (1-v^2)^{\frac{1}{7-p}} (1 + C v^2)^{\frac{1}{2}} T, 
\label{draggedbrane}
\end{equation}
where $C$ is given at (\ref{C}).
Note that the expression for $T_*$ 
differs from that given by \cite{Karch} using a 
Lorentz boost argument; the power of the Lorentz factor $(1-v^2)$ is 
different and there is the Lorentz noninvariant factor $(1 + Cv^2)^{1/2}$.
Since the brane configuration (\ref{branebending})
shows bending in the bulk in a way that has nontrivial dependence on the velocity $v$, 
the relation between $T$ and $T_*$ does not necessarily follow from a Lorentz boost argument alone.

There are important similarities and differences from the case driven by the applied electric field. 
If we assume the velocity $v$ is small
and expand the effective temperature in powers of $v$, we find
\begin{equation}
  T_* = T + \frac{1}{2} \left(C - \frac{2}{7-p}\right) v^2 T + O(v^4).
\end{equation}
Interestingly, the condition for $T_* < T$ at the order of $v^2$ is identical to that in the previous case
at the order of $E^2$. 

On the other hand, the location of the worldvolume horizon $r_*$ is proportional to that of the bulk horizon $r_0$, and 
the $T_*$ is proportional to $T$. This is in contrast to the previous case when $r_*$ and $T_*$ remain finite
even in the limit of $T \rightarrow 0$ provided we turn on $E$.

When the probe is a fundamental string, 
we find that $T_* = (1-v^2)^{\frac{1}{7-p}}$, namely the Lorentz noninvariant factor $(1+Cv^2)^{\frac{1}{2}}$ is absent.
In particular, $T_* = (1-v^2)^{1/4} T$ for $p=3$, reproducing the
result of \cite{Gubser:2006nz,CasalderreySolana:2007qw}. Note that $T_* < T$ in this case.

\bigskip
\noindent
{\emph{Fluctuation-Dissipation Relation}} ---
The emergence of the event horizon on the brane and the fact that the Hawking temperature is
the same for all degrees of freedom on the brane suggest that the effective 
temperature $T_*$ is a robust and universal feature of the class of nonequilibrium systems discussed here. 
In this section, we show that the fluctuation-dissipation relation also gives the same effective temperature
and that fluctuations around the steady states obey the Maxwell-Boltzmann distribution at $T_*$.

The fluctuation-dissipation relation is a property of an equilibrium system at temperature $T$,
and it can be stated in terms of Green's functions as
\begin{equation}
G^{\text{sym}}_{ij}(\omega)=-\coth(\omega/2T){\rm Im} G^{\text{R}}_{ij}(\omega),
\label{fdt}
\end{equation}
where $G^{\text{sym}}_{ij}(\omega)$ is the symmetrized Wightman function
defined as a Fourier transform $\int dt e^{-i\omega t}$ of $\langle {\cal O}_{i}(t) {\cal O}_{j}(0)+ {\cal O}_{j}(0) {\cal O}_{i}(t) \rangle/2$
for a set of operators ${\cal O}_i$ and we assume that their expectation values $\langle {\cal O}_i \rangle$
have been subtracted. $G^{\text{R}}_{ij}(\omega)$, on the other hand,  is the retarded Green's function
defined as a Fourier transform of
$-i\theta(t)\langle[\delta {\cal O}_{i}(t),\delta {\cal O}_{j}(0)]\rangle$.

In the limit $\omega \rightarrow 0$, this relation reduces to,
\begin{eqnarray}
\kappa_{ij}= 2T \eta_{ij},
\label{FDTT}
\end{eqnarray}
where $\kappa_{ij}=G^{\text{sym}}_{ij}|_{\omega=0}$ represents the noise, namely the strength of fluctuation, whereas $\eta_{ij}=-\lim_{\omega\to 0}{\rm Im} G^{\text{R}}_{ij}(\omega)/\omega$ represents the dissipation.
Holographically, the relation (\ref{fdt}) 
arises from the Schwinger-Keldysh formalism in AdS/CFT~\cite{Herzog:2002pc},
where the temperature $T$ defined by (\ref{FDTT}) is 
the Hawking temperature at the horizon. Note that $T$ in this case is independent of which operator one uses
to observe it.

The fluctuation-dissipation relation (\ref{fdt}) holds for the class of 
nonequilibrium systems studied in this paper, where
the temperature $T$ is replaced by 
the effective temperature $T_*$, as
\begin{equation}
G^{\text{sym}}_{ij}(\omega)=-\coth(\omega/2T_*){\rm Im} G^{\text{R}}_{ij}(\omega),
\label{NEfdt}
\end{equation}
as expected from the Schwinger-Keldysh formalism applied to degrees of freedom on the branes \cite{Gubser:2006nz,CasalderreySolana:2007qw}.
The effective temperature $T_*$ can be determined by the measurements of noise and dissipation via 
\begin{eqnarray}
\kappa_{ij}= 2T_* \eta_{ij}.
\label{FDTTnonlinear}
\end{eqnarray}
Let us examine implications of (\ref{FDTTnonlinear}) for the cases studied in the previous sections.

\medskip
\noindent
(1) Electric Field

For the case with the applied electric field, we take ${\cal O}_i$ to be the electric current density $J_i$.  
In the linear response regime, $\eta_{ij}$ is identified with the conductivity $\sigma_{ij}$ by the Kubo formula. 
The fluctuation-dissipation relation (\ref{FDTT}) then reproduces the Jonson-Nyquist noise, $\kappa_{ij} = 2T \sigma_{ij}$. 

When nonlinear effects in the applied electric field $E$ are relevant,
the differential conductivity $\sigma_{ij}^{\text{diff}}=\partial J_{i}/\partial E^{j}$ 
is different from  the standard conductivity defined by $J_i = \sigma_{ij} E^j$, and it is 
the former that is directly related to the retarded Green's function as
\begin{eqnarray}
-\left.\lim_{\omega\to 0}\frac{{\rm Im} G^{\text{R}}_{ij}(\omega)}{\omega}\right.
=\left.\frac{\partial J_{i}}{\partial E^{j}}\right.
=\left. \left(\sigma_{ij}+\frac{\partial \sigma_{ik}}{\partial E^{j}} E^{k}\right)\right. .
\:\:\:\:\:\:\:\:
\label{dif-con}
\end{eqnarray}
We have used our holographic models to check
the relation (\ref{dif-con}) explicitly by computing  $G^{\text{R}}_{ij}$
with the ingoing-wave boundary condition at the worldvolume horizon and comparing its imaginary 
part with the differential conductivity directly obtained from (\ref{J}).

The generalized fluctuation-dissipation relation (\ref{FDTTnonlinear}) then implies
\begin{eqnarray}
\kappa_{TT}=2 T_{*}\sigma_{TT},\:\:
\kappa_{LL}=2 T_{*}\left. \left(\sigma_{LL}+E^{L}\frac{\partial \sigma_{LL}}{\partial E^{L}}\right)\right. ,\:\:\:\:\:\:\:\:
\label{FDT-NESS-2}
\end{eqnarray}
where  $L$ and $T$ refer, respectively, to directions longitudinal and transverse to the applied electric field.
Though these equations show that current fluctuations are not isotropic, the effective temperature $T_{*}$
is isotropic by definition. 
The anisotropy originates from the nonlinearity of the differential conductivity. 
The transverse (TT) part of (\ref{FDT-NESS-2}) in the case of  $p=3$ and $q=2$ has previously been verified in \cite{Sonner}. 

\medskip
\noindent
(2) Drag Force

Here we will focus our attention to the case with $p=3, q=0$ and $n=0$. Fluctuation of the force acting on the dragged particle has been computed in \cite{CasalderreySolana:2007qw,Gubser:2006nz,Giecold:2009cg}, and the relationship between $T_{*}$ and fluctuation-dissipation relation corresponding to (\ref{FDTTnonlinear}) has been studied in \cite{Gursoy:2010aa}.

In \cite{Giecold:2009cg}, it was noted that momentum fluctuations in the transverse and longitudinal 
directions, denoted by $p_T$ and $p_L$ respectively, obey different distributions,
\begin{eqnarray}
P(p_T) &\simeq& \frac{1}{2\pi m \gamma T_*}\exp\left(- \frac{p_T^2}{2m \gamma T_*}\right), \nonumber \\
P(p_L) &\simeq& \frac{1}{\sqrt{2\pi m \gamma^3 T_*}}\exp\left(-\frac{p_L^2}{2m\gamma^3 T_*}\right),
\label{Maxwell}
\end{eqnarray}
where $\gamma = 1/\sqrt{1-v^2}$ and $m$ is the mass of the dragged object. 
This raised a question on whether one can define an effective temperature that applies to all degrees of freedom on the
brane. 

We would like to make a point that the energy distribution, which is
more directly related to temperature, is isotropic. 
Suppose the momentum fluctuates around  $P$ as $ P \rightarrow P +p$ and expand the kinetic energy as,
\begin{eqnarray}
{\cal E}&=&\sqrt{m^{2}+(P+p_{L})^{2}+p_{T}^{2}}
\nonumber \\
&=&m\gamma 
+\frac{\langle p_{T}^{2} \rangle}{2m\gamma}
+\frac{\langle p_{L}^{2} \rangle}{2m\gamma^{3}} + \cdots.
\label{drag_energy}
\end{eqnarray}
The Lorentz factors $1/\gamma$ and $1/\gamma^3$ that accompany $p_T^2$ and $p_L^2$ in the above are
exactly the same as those appear in the momentum distribution (\ref{Maxwell}). Thus,
both the transverse and the longitudinal momentum fluctuations obey the Maxwell-Boltzmann distribution at temperature $T_*$.

Our definition of the effective temperature is related to some of those discussed in the literature on nonequilibrium statistical 
physics. In \cite{effectiveT}, for example, the effective temperature is defined  by (17) and (21), 
i.e., the fluctuation-dissipation relation. These correspond to (\ref{NEfdt}) and (\ref{FDTTnonlinear})
of this paper. 

For our definition of $T_*$, it has been essential that the degrees of freedom on the brane is decoupled from those 
in the bulk. In particular, we have ignored back-reactions to the bulk degrees of freedom. Since the causal
structures on the brane and in the bulk are different---for example, the horizons are located differently at
$r_0$ and $r_*$---it is likely that coupling the degrees of freedom on the brane and in the bulk will 
modify the notion of the effective temperature $T_*$ on the brane.

\bigskip
\noindent
{\emph{Negative Excess Noise}} ---
Whereas the effective temperature $T_*$ is the same for all the degrees of freedom on the brane,
the fluctuation $\kappa_{ij}$ can depend on which degree of freedom we observe.  
For the current noises in (\ref{FDT-NESS-2}),
\begin{eqnarray}
\kappa(E)=\kappa(E=0)\left[1+\frac{1}{2}\widetilde{C}\frac{E^{2}}{(c_{0}T)^{\frac{14-2p}{5-p}}}+O(E^{4})\right],
\end{eqnarray}
where
\begin{eqnarray}
\widetilde{C}=C-\frac{2}{7-p}+\alpha(C-1),
\end{eqnarray}
with $\alpha=3$ for the longitudinal fluctuations and $\alpha=1$ for the transverse fluctuations. If $\widetilde{C}<0$, the current noise at the order of $E^{2}$ is smaller than that at $E=0$, namely, the noise is reduced by driving the system by the electric field. 
The reduction of noise may be counterintuitive, but it is not forbidden and is known as negative excess noise \cite{NEN}. Our models provide explicit examples of realization of this curious phenomenon.
This occurs if $T_{*}<T$ for $p<5$,
for example. However, we find that $T_{*}<T$ is not a necessary condition, e.g., it also happens when $q=1$ and $p=3$ where $T_{*}>T$. 

For the dragged branes studied in this paper, the excess noise is negative to the order of $v^{2}$ 
if $C-\frac{2}{7-p}+\alpha(C+1) < 0$, where $\alpha=3$ ($\alpha=1$) for the longitudinal (transverse) fluctuations. 
For $p < 5$, this occurs only for the transverse mode in the case of $(p, q, n) = (4, 0 , 1)$, 
corresponding to a D2 brane in a heat bath of D4 branes (even in this case, the longitudinal mode has positive
excess noise).  In all other cases, excess noises are positive for $p < 5$ even when  $T_{*}<T$.


\bigskip

\noindent
{\emph{Acknowledgments}.--- We thank S.~S.~Gubser, C.~P.~Herzog, A.~Karch, E.~Kiritsis, K.~Kobayashi, H.~Liu, 
S.~Sasa, T.~Takayanagi, D.~Teaney and Y.~Utsumi for discussions and comments. 
The work of H.O. is supported in part by U.S. DOE Grant No.
DE-FG03-92-ER40701,
the Simons Foundation, JSPS Grant-in-Aid for Scientific Research 
C-23540285, and the WPI Initiative of MEXT of Japan. He also thanks the
hospitality of the Aspen Center for Physics and the National Science Foundation, which supports the Center under Grant No. PHY-1066293, 
and of the Simons Center for Geometry and Physics.
The work of S.N. was supported in part by the Grant-in-Aid for Scientific Research on Innovative Areas No. 2104, and Grant-in-Aid for Challenging Exploratory Research $\sharp$ 23654132.

\end{document}